\documentclass{emulateapj}

\shorttitle{Chemical segregation toward the AFGL2591 hot core}
\shortauthors{I. Jim\'enez-Serra et al.}

\begin{document}

\title{Chemical segregation toward massive hot cores: The AFGL2591 star forming region}

\author{I. Jim\'{e}nez-Serra\altaffilmark{1}, Q. Zhang\altaffilmark{1}, S. Viti\altaffilmark{2}, J. Mart\'{\i}n-Pintado\altaffilmark{3} and W.-J. de Wit\altaffilmark{4}}

\altaffiltext{1}{Harvard-Smithsonian Center for Astrophysics, 
60 Garden St., Cambridge, MA 02138, USA; ijimenez-serra@cfa.harvard.edu; qzhang@cfa.harvard.edu}

\altaffiltext{2}{Department of Physics \& Astronomy, University College London, Gower Place, London WC1E 6BT, United Kingdom; sv@star.ucl.ac.uk}

\altaffiltext{3}{Centro de Astrobiolog\'{\i}a (CSIC/INTA),
Ctra. de Torrej\'on a Ajalvir km 4,
E-28850 Torrej\'on de Ardoz, Madrid, Spain; 
jmartin@cab.inta-csic.es}

\altaffiltext{4}{European Southern Observatory, Alonso de Cordova 3107, Vitacura, Santiago, Chile; wdewit@eso.org}

\begin{abstract}

We present high angular resolution observations (0.5$"$$\times$0.3$"$) carried out with the Submillimeter Array (SMA) toward the AFGL2591 high-mass star forming region. Our SMA images reveal a clear chemical segregation within the AFGL2591 VLA 3 hot core, where different molecular species (Type I, II and III) appear distributed in three concentric shells. This is the first time that such a chemical segregation is ever reported at linear scales $\leq$3000$\,$AU within a hot core. While Type I species (H$_2$S and $^{13}$CS) peak at the AFGL2591 VLA 3 protostar, Type II molecules (HC$_3$N, OCS, SO and SO$_2$) show a double-peaked structure circumventing the continuum peak. Type III species, represented by CH$_3$OH, form a ring-like structure surrounding the continuum emission. The excitation temperatures of SO$_2$, HC$_3$N and CH$_3$OH (185$\pm$11$\,$K, 150$\pm$20$\,$K and 124$\pm$12$\,$K, respectively) show a temperature gradient within the AFGL2591 VLA 3 envelope, consistent with previous observations and modeling of the source. By combining the H$_2$S, SO$_2$ and CH$_3$OH images, representative of the three concentric shells, we find that the global kinematics of the molecular gas follow Keplerian-like rotation around a 40$\,$M$_\odot$-star. The chemical segregation observed toward AFGL2591 VLA 3 is explained by the combination of molecular UV photo-dissociation and a high-temperature ($\sim$1000 K) gas-phase chemistry within the low extinction innermost region in the AFGL2591 VLA 3 hot core.

\end{abstract}


\section{Introduction}

Hot cores are hot ($\sim$200$\,$K), compact ($\leq$0.1$\,$pc) and dense ($\geq$10$^6$$\,$cm$^{-3}$) condensations, which represent one of the earliest stages of massive star formation \citep[e.g.][]{gar99}. These objects are chemically very rich, and show large enhancements of S-bearing species such as H$_2$S, OCS or SO$_2$, and of complex organic molecules (COMs) such as H$_2$CO, CH$_3$OH or CH$_3$CN \citep[][]{bla87,com05,hat98b}. These enhancements are attributed to the thermal evaporation of the mantles of dust grains, followed by rapid ion-neutral gas-phase reactions \citep{cha92,cha97,hat98a,viti04,wake04}.    

Toward hot cores, the warming up of the gas in the envelope by the central protostar, generates a temperature gradient as a function of radius that largely affects the chemical composition of the grain mantles and of the molecular gas within these cores. As a consequence, a chemical segregation is expected to occur \citep[][]{van98,nom04}. Indirect evidence of this molecular segregation has been found in the form of abundance discontinuities for COMs toward hot cores \citep{van99,van00b}, as well as toward low-mass warm cores \citep[or {\it hot corinos}; see][]{cec00,mar04,awad10}. However, direct imaging of the chemical segregation at the inner regions of hot molecular cores, still remains to be reported.

The AFGL2591 high-mass star forming region is located in the Cygnus X complex at a distance of $\sim$3$\,$kpc. This distance, which was substantially underestimated in the past, has been derived recently by using trigonometric parallax of the 22.2$\,$GHz H$_2$O masers detected toward this region \citep[][]{rygl12}. The revised mass of the large-scale AFGL2591 clump is 2$\times$10$^4$$\,$M$_\odot$, and the measured total IR luminosity is $\sim$2$\times$10$^5$$\,$L$_\odot$ \citep{san12}. AFGL2591 hosts a cluster of young stellar objects (YSOs) over scales of 0.1$\,$pc \citep{cam84,tri03,van05}, where several radio continuum sources are detected \citep[see VLA 1, VLA 2 and VLA 3 in][]{tri03}. \citet{san12} has recently reported the detection of another YSO (the NE source) at $\sim$0.4$"$ (1300$\,$AU) north the AFGL2591 VLA 3 object.

From all the YSOs detected in the AFGL2591 star forming region, AFGL2591 VLA 3 is believed to be the youngest \citep[dynamical age of $\sim$2$\times$10$^4$$\,$yr;][]{doty02,sta05} and most massive source in the cluster \citep[estimated mass of $\sim$38$\,$M$_\odot$;][]{san12}. AFGL2591 VLA 3 is very bright in the mid-IR \citep[][]{dewit09}, and is responsible for most of the IR luminosity measured toward this region \citep{san12}. AFGL2591 VLA 3 is also known to power a large-scale east-west outflow detected in CO and near-IR emission \citep[P.A.$\sim$280$^\circ$;][]{mit92,prei03}, whose axis lies close to the direction of the line-of-sight \citep[][]{has95,van99}. The blue-shifted lobe of the outflow is detected toward the west of AFGL2591 VLA 3, while the red-shifted outflowing gas is found toward the east \citep{mit92}. The AFGL2591 VLA 3 molecular envelope is seen almost pole-on, and rotates in the counter-clockwise direction \citep{van06}. \citet{van99} have proposed that the AFGL2591 VLA 3 molecular envelope has two different physical regimes with an inner and hotter core and an outer and cooler envelope. This makes the AFGL2591 VLA 3 hot core an excellent candidate to probe for chemical segregation.

In this paper, we report the first detection of a clear chemical segregation at linear scales $\leq$3000$\,$AU within the hot core around the AFGL2591 VLA 3 high-mass protostar. The comparison of the SMA images with gas-grain chemical models shows that the observed chemical segregation is due to i) a strong UV radiation field leading to molecular photo-dissociation; and ii) a high-temperature gas-phase chemistry within the low-extinction innermost region in this core. The paper is organized as follows. The SMA observations are reported in Section$\,$\ref{obs}. The images of the continuum and molecular line emission toward AFGL2591 VLA 3 are presented in Section$\,$\ref{res}. The derived excitation temperatures, column densities and abundances of the molecular species measured toward this source are given in Section$\,$\ref{exc}. And the modelling of the chemical segregation detected toward this object is presented in Section$\,$\ref{chem}. In Section$\,$\ref{conc}, we summarize our conclusions.

\section{Observations}
\label{obs}

\begin{figure}
\begin{center}
\includegraphics[angle=0,width=0.4\textwidth]{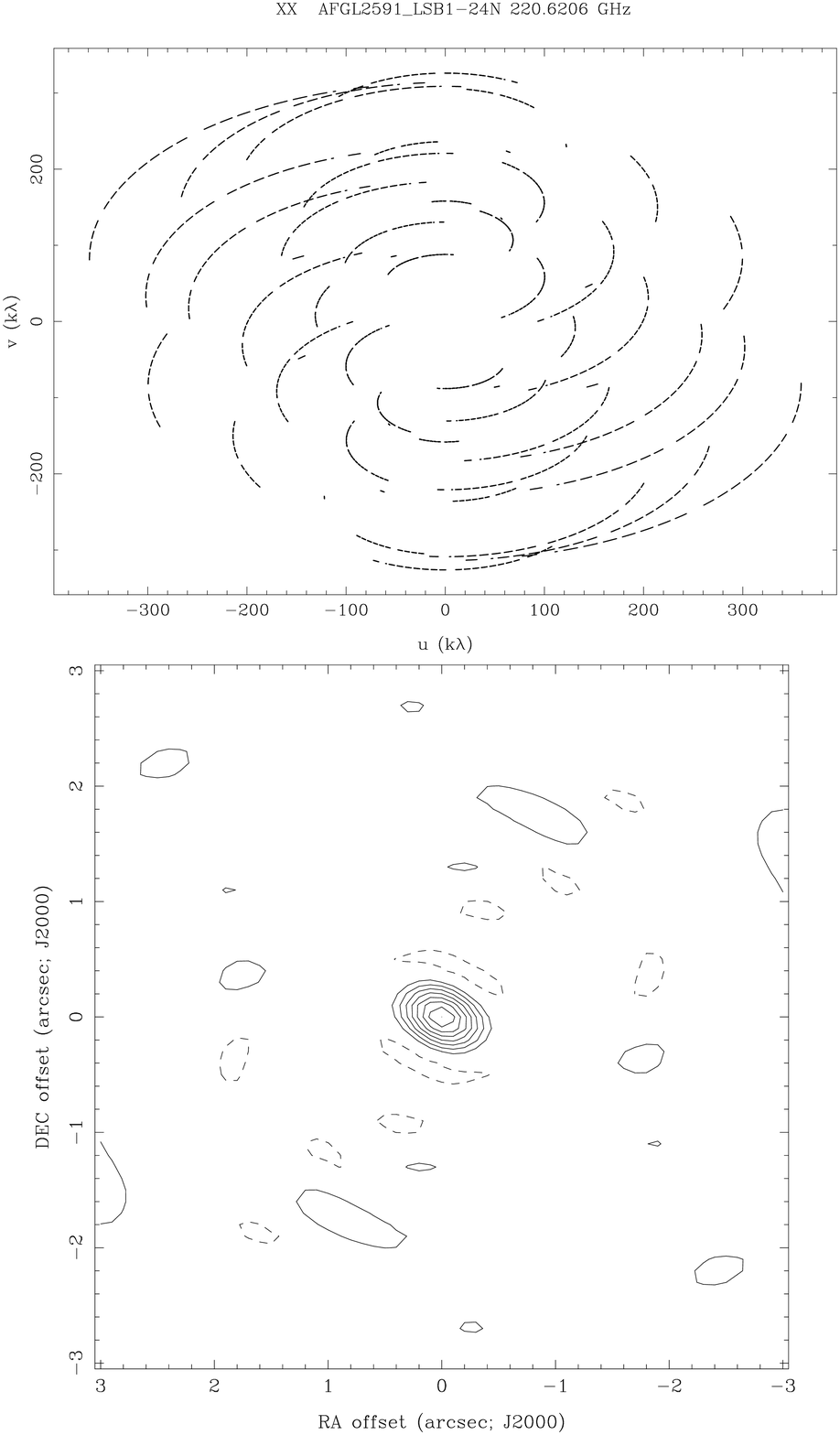}
\caption{UV coverage (upper panel) and dirty beam (lower panel) of our observations taken with the SMA in VEX configuration. The synthesized beam is 0.55$"$$\times$0.34$"$ (Section$\,$\ref{obs}).}
\label{f1}
\end{center}
\end{figure}

Observations of the AFGL2591 star forming region were carried out with the SMA\footnote{The Submillimeter Array is a joint project between the Smithsonian Astrophysical Observatory and the Academia Sinica Institute of Astronomy and Astrophysics, and is funded by the Smithsonian Institution and the Academia Sinica.} in the very extended configuration (VEX) on 2010 February 15. The size of the synthesized beam of our VEX observations was 0.55$"$$\times$0.34$"$ (see Figure$\,$\ref{f1} for the UV coverage and dirty beam of the VEX data). 
The phase center of the observations was set at $\alpha(J2000)$=20$^{h}$29$^{m}$24.90$^s$, $\delta(J2000)$=+40$^{\circ}$11$'$20.3$''$. The zenith opacity at 225$\,$GHz was 0.06 and the double sideband system temperatures were typically 200-280$\,$K. The receivers were tuned to an LO frequency of 218.75$\,$GHz, and the correlator provided a uniform spectral resolution of 0.8$\,$MHz (i.e. $\sim$1.1$\,$km$\,$s$^{-1}$). The radial velocity of the source was set at -5.5$\,$km$\,$s$^{-1}$ \citep{van99}. 
We used 3C273 (12.5$\,$Jy) as bandpass calibrator, and MWC349A (1.7$\,$Jy) as flux and gain calibrator. Calibration of the raw data was carried out within the IDL MIR software package, and continuum subtraction, imaging and deconvolution was done within MIRIAD. 

In order to quantitatively measure the effect of missing flux in our VEX data (see Section$\,$\ref{uv}), we have also used unpublished data obtained by us with the SMA in Subcompact configuration (beam of 6.7$"$$\times$5.6$"$). The central coordinates and spectral resolution of the SMA Subcompact images are the same as those of the VEX observations. However, the frequency coverage was slightly different from that of the VEX data, and the H$_2$S (2$_{2,0}$$\rightarrow$2$_{1,1}$) line was not observed. A detailed analysis of the large-scale structure of the continuum and molecular line emission toward AFGL2591 VLA 3 will be presented elsewhere (Jim\'enez-Serra et al. in preparation). In the present paper, we only use the Subcompact data to demonstrate that the chemical segregation detected toward this source is not an effect of missing short spacings in the VEX observations.

%FIG1**************************************
%******************************************

\section{Results}
\label{res}

\subsection{Dust continuum emission and $^{12}$CO outflowing gas}
\label{co}

\begin{figure}
\begin{center}
\includegraphics[angle=270,width=0.45\textwidth]{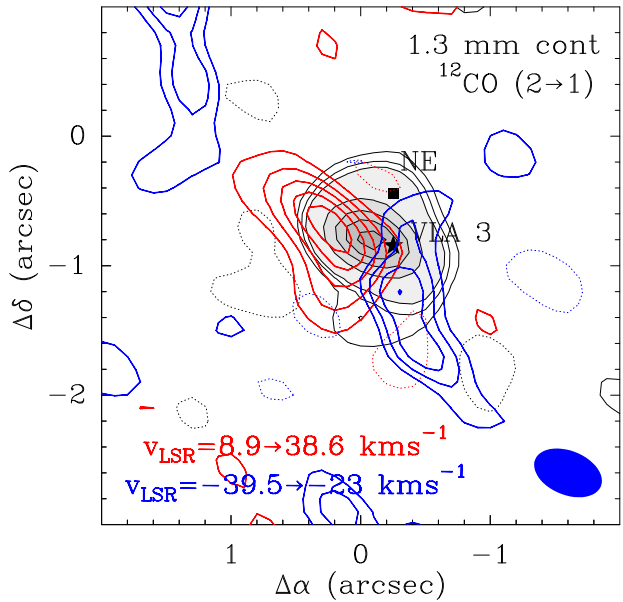}
\caption{Integrated intensity maps of the $^{12}$CO $J$=2$\rightarrow$1 emission detected toward AFGL2591 VLA 3 from -39.5 to -23$\,$km$\,$s$^{-1}$ (blue contours) and from 8.9 to 38.6$\,$km$\,$s$^{-1}$ (red contours), superimposed on the 1.3$\,$mm dust continuum map (gray scale and black contours). The contour levels for the 1.3$\,$mm emission are 2.4 (3$\sigma$), 4.0, 5.6, 17.6, 29.6, 41.6 and 53.6$\,$mJy/beam. For $^{12}$CO, the first contour and step levels are 0.30 (2$\sigma$) and 0.15$\,$Jy/beam$\,$km$\,$s$^{-1}$ for the blue-shifted lobe, and 0.36 (2$\sigma$) and 0.54$\,$Jy/beam$\,$km$\,$s$^{-1}$ for the red-shifted lobe. Negative contours correspond to the 3$\sigma$ level. Filled star and filled square show the location of the VLA 3 and NE sources \citep{tri03,san12}. The beam is shown at the lower right corner.}
\label{f2}
\end{center}
\end{figure}

In Figure$\,$\ref{f2}, we show the $^{12}$CO images for the red- and blue-shifted gas detected toward AFGL2591 VLA 3 (red and blue contours, respectively), superimposed on the 1.3$\,$mm dust continuum emission (gray scale and black contours). The continuum emission (size of $\sim$0.6$"$$\times$0.45$"$) is associated with the brightest mid-IR source in the region \citep{dewit09}, and peaks at $\alpha(J2000)$=20$^{h}$29$^{m}$24.889$^s$, $\delta(J2000)$=+40$^{\circ}$11$'$19.51$''$. This position is very similar to that reported by \citet{ben07} for the continuum source seen at 348$\,$GHz in their SMA images. The derived integrated 1.3$\,$mm continuum flux ($\sim$100$\,$mJy) is smaller than previously reported \citep[$\sim$190$\,$mJy;][]{van06}, because the PdBI observations (beam of $\sim$0.9$"$) were more sensitive to extended emission. The VLA 3 source detected at 3.6$\,$cm \citep[][]{tri03} is located at $\sim$0.2$"$ east of the 1.3$\,$mm dust continuum peak, possibly linked to the blue-shifted part of the high-velocity jet \citep{san12}. The fainter 1.3$\,$mm continuum component toward the south of AFGL2591 VLA 3 could be associated with shocked gas at the walls of the outflow cavity \citep[see][]{mau10}. 

Toward the north of VLA 3 (see Figure$\,$\ref{f2}), the 1.3$\,$mm continuum emission shows an elongation that could be associated with the NE source reported by \citet{san12}. This source, which is not detected at centimeter wavelengths \citep{tri03}, has likely generated the double bow-shock structure seen in H$_2$O masers at 0.4$"$ north VLA 3. As shown in Sections$\,$\ref{mole} and \ref{kin}, the NE source does not have any clear counterpart in molecular line emission. The kinematics of the molecular gas toward AFGL2591 VLA 3 seem to be dominated by the kinematics of the VLA 3 molecular envelope, although the angular resolution of our observations may not be sufficient to resolve any molecular feature arising from the NE source.

From Figure$\,$\ref{f2}, we find that the $^{12}$CO emission forms a bi-conical structure aligned in the east-west direction and centered at the 1.3$\,$mm VLA 3 source. The kinematics of this structure are consistent with those of the large-scale $^{12}$CO outflow \citep[blue-shifted gas toward the west and red-shifted emission toward the east; see][]{mit92}. In addition, the $^{12}$CO blue lobe detected in our VEX images matches the morphology of the bright near-IR loop detected in the K-band \citep{prei03}. The opening angle of the bi-conical structure detected with the SMA is $\sim$90$^\circ$-100$^\circ$, similar to that measured from H$_2$O masers \citep[110$^\circ$;][]{san12} or from near-IR emission \citep[$>$100$^\circ$;][]{prei03}.  

%FIG2**************************************
%******************************************

\subsection{Differences in the morphology of the molecular line emission toward AFGL2591 VLA 3}
\label{mole}

\begin{figure*}
\begin{center}
\includegraphics[angle=270,width=0.95\textwidth]{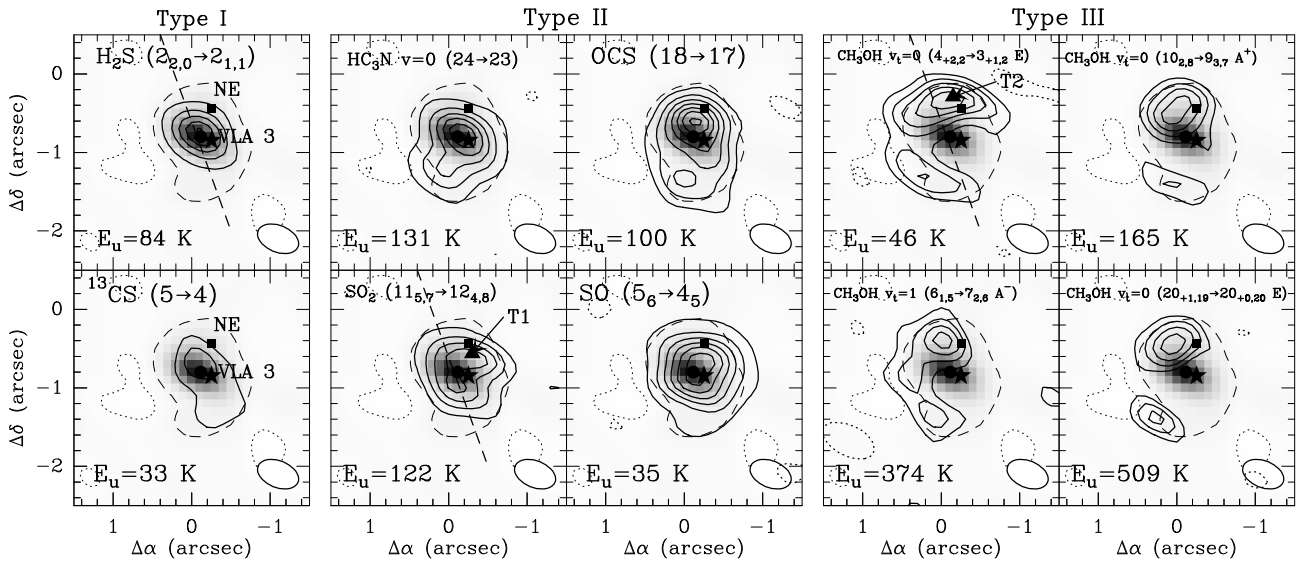}
\caption{Integrated intensity images of several molecular species from  -8.7 to -2.1$\,$km$\,$s$^{-1}$ (thick contours), superimposed on the 1.3$\,$mm continuum map (color scale and dashed contour). Filled star, filled square and filled circle show, respectively, the location of the VLA 3 source \citep{tri03}, the NE source \citep{san12}, and the main 1.3$\,$mm dust continuum peak detected with the SMA (Section$\,$\ref{co}). The molecular species, transition and energy of the upper level of every map are shown in every panel. The first contour and step levels are: 0.27 (3$\sigma$) and 0.45$\,$Jy/beam$\,$km$\,$s$^{-1}$ for H$_2$S (2$_{2,0}$$\rightarrow$2$_{1,1}$), 0.24 (3$\sigma$) and 0.24$\,$Jy/beam$\,$km$\,$s$^{-1}$ for $^{13}$CS (5$\rightarrow$4), 0.21 (3$\sigma$) and 0.35$\,$Jy/beam$\,$km$\,$s$^{-1}$ for HC$_3$N $v$=0 (24$\rightarrow$23), 0.21 (3$\sigma$) and 0.21$\,$Jy/beam$\,$km$\,$s$^{-1}$ for OCS (18$\rightarrow$17), 0.21 (3$\sigma$) and 0.28$\,$Jy/beam$\,$km$\,$s$^{-1}$ for SO$_2$ (11$_{5,7}$$\rightarrow$12$_{4,8}$), 0.36 (3$\sigma$) and 0.60$\,$Jy/beam$\,$km$\,$s$^{-1}$ for SO (5$_6$$\rightarrow$4$_5$), 0.21 (3$\sigma$) and 0.21$\,$Jy/beam$\,$km$\,$s$^{-1}$ for CH$_3$OH $v_t$=0 (4$_{+2,2}$$\rightarrow$3$_{+1,2}$ E), 
0.18 (2$\sigma$) and 0.18$\,$Jy/beam$\,$km$\,$s$^{-1}$ for CH$_3$OH $v_t$=0 (10$_{2,8}$$\rightarrow$9$_{3,7}$ A$^+$), 0.15 (2$\sigma$) and 0.15$\,$Jy/beam$\,$km$\,$s$^{-1}$ for CH$_3$OH $v_t$=1 (6$_{1,5}$$\rightarrow$7$_{2,6}$ A$^-$), and 0.14 (2$\sigma$) and 0.07$\,$Jy/beam$\,$km$\,$s$^{-1}$ for CH$_3$OH $v_t$=0 (20$_{+1,19}$$\rightarrow$20$_{+0,20}$ E). Negative contours correspond to the 3$\sigma$ level. Dashed lines show the direction of the PV cuts reported in Figure$\,$\ref{f5}. T1 and T2 are the brightest emission peaks seen in SO$_2$ and CH$_3$OH, where T$_{ex}$ has been estimated (Section$\,$\ref{exc} and Figure$\,$\ref{f7}). Beams are shown at the lower right corner in every panel.}
\label{f3}
\end{center}
\end{figure*}

In Table$\,$\ref{tab1}, we report the molecular rotational lines measured within the 8$\,$GHz passband of the SMA, and Figure$\,$\ref{f3} presents the integrated intensity maps for some of these lines superimposed on the 1.3$\,$mm continuum image of the AFGL2591 VLA 3 source. From Figure$\,$\ref{f3}, we find that the observed molecular species show different spatial distributions in a clear chemical segregation. According to the observed morphology, we classify these molecules into Type I, II and III species (see Figure$\,$\ref{f3}). While Type I species (H$_2$S) have a compact structure peaking at the AFGL2591 VLA 3 continuum emission (left panels), Type II molecules (HC$_3$N, OCS and SO$_2$) show a double-peaked morphology that circumvents the continuum peak (middle panels). For Type III species (represented by CH$_3$OH; right panels), the emission is distributed in a ring-like structure surrounding the continuum core. 

%TABLE1------------------------------------------------------------------
\begin{deluxetable}{lccc}
\tabletypesize{\scriptsize}
\tablecaption{Molecular line transitions observed toward AFGL2591 VLA 3 with the SMA.}
\tablewidth{0pt}

\startdata

%\hline \hline

& \multicolumn{2}{c}{TABLE 1} & \\ \hline \hline
Molecule & Transition & Frequency (MHz) & $E_u$ (K) \\ \hline

H$_2$S & 2$_{2,0}$$\rightarrow$2$_{1,1}$ & 216710.44 & 84 \\ 
$^{13}$CS & 5$\rightarrow$4 & 231220.99 & 33 \\ 
SO & 5$_6$$\rightarrow$4$_5$ & 219949.44 & 35 \\ 
OCS & 18$\rightarrow$17 & 218903.36 & 100 \\
& 19$\rightarrow$18 & 231060.98 & 111 \\ 
HC$_3$N & $v$=0 24$\rightarrow$23 & 218324.72 & 131 \\
& $v_6$=1$_e$ 24$\rightarrow$23 & 218682.56 & 850 \\
& $v_6$=1$_f$ 24$\rightarrow$23 & 218854.39 & 850 \\
& $v_7$=1$_e$ 24$\rightarrow$23 & 218860.80 & 453 \\
& $v_7$=1$_f$ 24$\rightarrow$23 & 219173.76 & 453 \\
SO$_2$ & 22$_{7,15}$$\rightarrow$23$_{6,18}$ & 219275.98 & 353 \\
& 22$_{2,20}$$\rightarrow$22$_{1,21}$ & 219465.55 & 994 \\
& 16$_{3,13}$$\rightarrow$16$_{2,14}$ & 220165.26 & 893 \\
& 11$_{5,7}$$\rightarrow$12$_{4,8}$ & 229347.63 & 122 \\
& 13$_{2,12}$$\rightarrow$13$_{1,13}$ & 229545.30 & 838 \\
& 37$_{10,28}$$\rightarrow$38$_{9,29}$ & 230965.25 & 892 \\
& 14$_{3,11}$$\rightarrow$14$_{2,12}$ & 231980.53 & 865 \\
CH$_3$OH & $v_t$=0 5$_{+1,4}$$\rightarrow$4$_{+2,2}$ $E$ & 216945.52 & 56 \\
& $v_t$=1 6$_{1,5}$$\rightarrow$7$_{2,6}$ $A^-$ & 217299.20 & 374 \\
& $v_t$=0 20$_{+1,19}$$\rightarrow$20$_{+0,20}$ $E$ & 217886.50 & 509 \\
& $v_t$=0 4$_{+2,2}$$\rightarrow$3$_{+1,2}$ $E$ & 218440.06 & 46 \\
& $v_t$=0 8$_{+0,8}$$\rightarrow$7$_{+1,6}$ $E$ & 220078.56 & 97 \\
& $v_t$=0 15$_{+4,11}$$\rightarrow$16$_{+3,13}$ $E$ & 229589.06 & 375 \\
& $v_t$=0 8$_{-1,8}$$\rightarrow$7$_{+0,7}$ $E$ & 229758.76 & 89 \\
& $v_t$=0 19$_{5,15}$$\rightarrow$20$_{4,16}$ $A^+$ & 229864.12 & 579 \\
& $v_t$=0 19$_{5,14}$$\rightarrow$20$_{4,17}$ $A^-$ & 229939.10 & 579 \\
& $v_t$=0 3$_{-2,2}$$\rightarrow$4$_{-1,4}$ $E$ & 230027.05 & 40 \\
& $v_t$=0 10$_{2,9}$$\rightarrow$9$_{3,6}$ $A^-$ & 231281.11 & 165 \\
& $v_t$=0 10$_{2,8}$$\rightarrow$9$_{3,7}$ $A^+$ & 232418.52 & 165 \\ \hline

\enddata

\label{tab1}
\end{deluxetable}
%------------------------------------------------------------------

%FIG3**************************************
%******************************************

Two species, namely $^{13}$CS and SO, have been tentatively classified as Type I and Type II species, respectively. For $^{13}$CS, although the brightest emission is detected toward the continuum peak, this molecule also presents a secondary peak toward the south-west of AFGL2591 VLA 3. This peak does not coincide with any of the sources detected in the region or with the molecular emission peaks from Type II or III species. The secondary peak could be associated with outflowing gas. However, its kinematics (at red-shifted velocities) are not consistent with those of the outflow toward this position. As shown in Section$\,$\ref{chem}, CS is also expected to be relatively abundant toward the regions where Type II species are detected, suggesting that the CS emission could be more extended than detected with the rarer $^{13}$CS isotopologue. 

For SO, the double-peaked morphology in the SO integrated intensity map of Figure$\,$\ref{f3}, is not as clearly seen as for HC$_3$N, OCS or SO$_2$. However, a closer inspection of the Position-Velocity diagram of the SO emission reveals a similar kinematic structure to that of e.g. SO$_2$ (see Figure$\,$\ref{f5} and below). Nevertheless, we cannot rule out the possibility that the emission from the low-excitation SO (5$_6$$\rightarrow$4$_5$) transition is affected by large optical depth effects and/or it is part of the base of the AFGL2591 VLA 3 outflow. A significant amount of SO could also be present within the same region where Type I species arise (i.e. H$_2$S and $^{13}$CS), as suggested by the modeling of the chemistry of the AFGL2591 VLA 3 source (Section$\,$\ref{chem}). 

We note that the double-peaked structure of Type II species such as SO$_2$ or OCS, and the ring-like morphology of CH$_3$OH, are not likely produced by excitation or optical depth effects since higher-excitation, optically thin transitions such as SO$_2$ (22$_{7,15}$$\rightarrow$23$_{6,18}$) ($E_u$=353$\,$K) or CH$_3$OH $v_t$=1 (6$_{1,5}$$\rightarrow$7$_{2,6}$ A$^-$) and CH$_3$OH $v_t$=0 (20$_{+1,19}$$\rightarrow$20$_{+0,20}$ E) ($E_u$=374$\,$K and 509$\,$K, respectively) present similar spatial distribution and kinematic structure to those of the lower-excitation lines (see Figure$\,$\ref{f3}, Figure$\,$\ref{f5} and Section$\,$\ref{kin} below). This behavior clearly contrasts with that observed toward the G29.96-0.02 hot core \citep{beu07}, where the CH$_3$OH high-excitation, optically thin lines peak toward the hot core while the low-excitation, optically thick transitions show a double-peaked morphology surrounding the core, as expected from large optical depths and if the CH$_3$OH gas were evenly distributed across the hot core. Although chemical differentiation has been reported toward G29.96-0.02 \citep[][]{beu07}, this differentiation is found across proto-cluster scales over $\sim$15000$\,$AU, i.e. a factor of 5 larger than the linear scales over which the chemical segregation toward AFGL2591 VLA 3 has been detected ($\leq$3000$\,$AU). Therefore, the chemical segregation toward AFGL2591 VLA 3 is likely related to the physical structure and physical processes taking place in the molecular envelope around the VLA 3 source.

One may also consider that the two main peaks seen in Type II and III species could be associated with two independent sources coincident with the bright 1.3$\,$mm continuum source toward AFGL2591 VLA 3 and the weaker secondary peak detected toward the south of this source (Figure$\,$\ref{f3}). However, species such as SO$_2$, OCS or CH$_3$OH show emission peaks which are located more than one beam away from these sources. Furthermore, the global kinematics of the molecular gas toward AFGL2591 VLA 3 fit well with Keplerian-like rotation when all molecular emission across the envelope is considered (see Figures$\,$\ref{f4} and \ref{f5}, and Section$\,$\ref{kin}), suggesting that this emission likely arises from the same physical entity rather than being associated with two different sources.

\subsection{Kinematics of the molecular gas toward AFGL2591 VLA 3}
\label{kin}

%FIG4**************************************
\begin{figure*}
\begin{center}
\includegraphics[angle=270,width=1.0\textwidth]{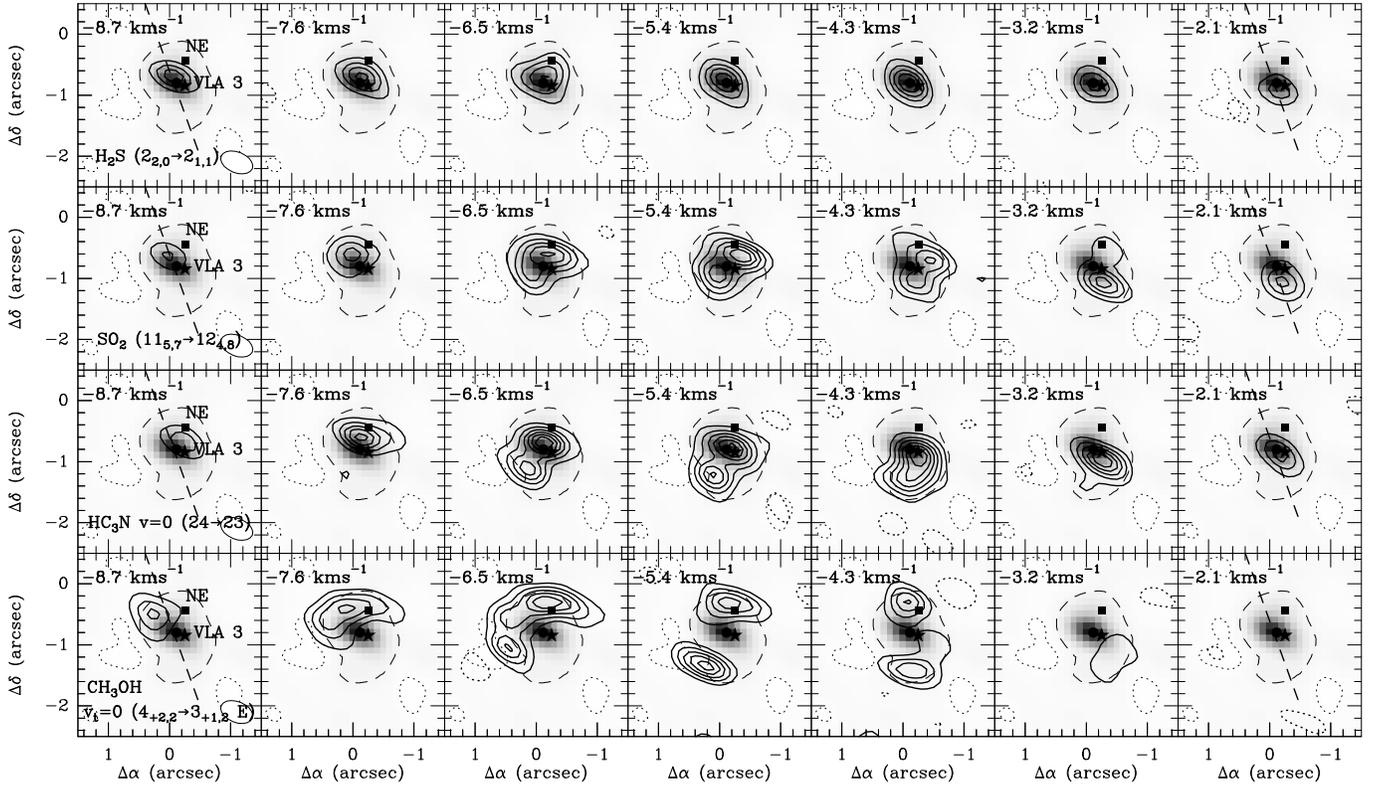}
\caption{Channel maps of the molecular emission of several lines of H$_2$S (Type I species), SO$_2$ and HC$_3$N (Type II), and CH$_3$OH (Type III), observed toward AFGL2591 VLA 3 (solid contours). These maps are superimposed on the 1.3$\,$mm continuum emission detected toward this source (in colour; dashed contour as in Figure$\,$\ref{f3}). The first contour and step levels of the molecular emission are, respectively, 0.075 (3$\sigma$) and 0.075$\,$Jy/beam for H$_2$S (2$_{1,0}$$\rightarrow$2$_{1,1}$), 0.06 (3$\sigma$) and 0.06$\,$Jy/beam for SO$_2$ (11$_{5,7}$$\rightarrow$12$_{4,8}$), 0.06 (3$\sigma$) and 0.06$\,$Jy/beam for HC$_3$N $v$=0 (24$\rightarrow$23), and 0.06 (3$\sigma$) and 0.06$\,$Jy/beam for CH$_3$OH $v_t$=0 (4$_{+2,2}$$\rightarrow$3$_{+1,2}$ E). The radial velocities of the channel maps are shown at the upper left corner in every panel. The central radial velocity of the AFGL2591 source is $v_{LSR}$=-5.5$\,$km$\,$s$^{-1}$ \citep{van99}. Dashed lines show the direction of the PV cuts reported in Figure$\,$\ref{f5}. Symbols are as in Figure$\,$\ref{f3}. The beam size is shown at the lower right corner in the first panels from the left.}
\label{f4}
\end{center}
\end{figure*}
%******************************************

Figure$\,$\ref{f4} presents the channel maps of some representative  lines of Type I (H$_2$S), Type II (SO$_2$ and HC$_3$N) and Type III (CH$_3$OH) species. These maps are superimposed on the 1.3$\,$mm continuum emission of the AFGL2591 VLA 3 source. From these maps, we find that the molecular gas toward VLA 3 shows a velocity gradient from the north-east to the south-west of this source, from blue- to red-shifted velocities. A similar behavior was reported for HDO and H$_2^{18}$O by \citet{van06}, and contrasts with that of the outflowing gas (Section$\,$\ref{co}). Since the outflow axis lies almost along the line-of-sight \citep{van99}, the kinematics of the molecular gas toward AFGL2591 VLA 3 are consistent with counter-clockwise rotation \citep{van06}. We note that the channel maps do not show any clear molecular feature arising from the recently reported source NE \citep{san12}.

\begin{figure*}
\begin{center}
\includegraphics[angle=270,width=0.85\textwidth]{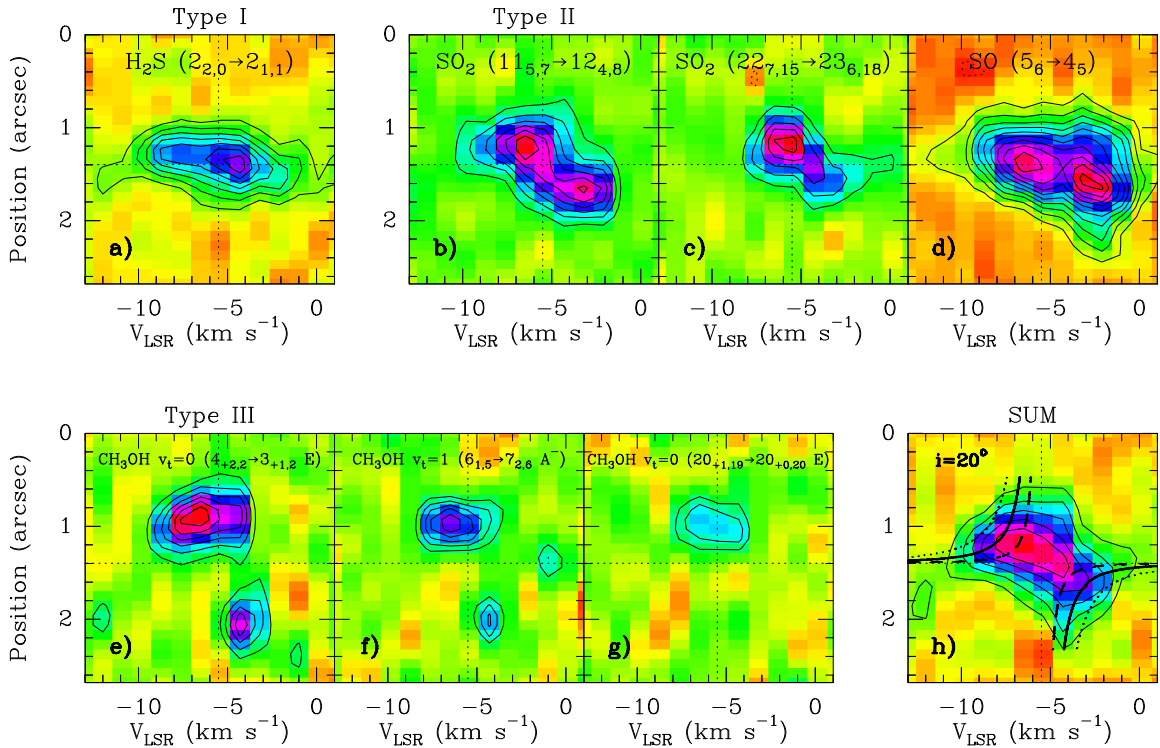}
\caption{PV diagrams of several lines of H$_2$S, SO$_2$, SO and CH$_3$OH observed toward AFGL2591 VLA 3 (panels from {\bf a} to {\bf g}). The first step and contour levels are 50 (2$\sigma$) and 50$\,$mJy/beam for H$_2$S (2$_{2,0}$$\rightarrow$2$_{1,1}$), 40 (2$\sigma$) and 40$\,$mJy/beam for SO$_2$ (11$_{5,7}$$\rightarrow$12$_{4,8}$), 60 (3$\sigma$) and 60$\,$mJy/beam for SO$_2$ (22$_{7,15}$$\rightarrow$23$_{6,18}$), 60 (3$\sigma$) and 60$\,$mJy/beam for SO (5$_6$$\rightarrow$4$_5$), 40 (2$\sigma$) and 40$\,$mJy/beam for CH$_3$OH $v_t$=0 (4$_{+2,2}$$\rightarrow$3$_{+1,2}$ E), 50 (2$\sigma$) and 25$\,$mJy/beam for CH$_3$OH $v_t$=1 (6$_{1,5}$$\rightarrow$7$_{2,6}$ A$^-$), and 50 (2$\sigma$) and 25$\,$mJy/beam for CH$_3$OH $v_t$=0 (20$_{+1,19}$$\rightarrow$20$_{+0,20}$ E). Offsets (in arcseconds) are calculated along the PV cuts shown in Figures$\,$\ref{f3} and \ref{f4} (P.A.$\sim$20$^\circ$) from the northeast to the southwest of AFGL2591 VLA 3. Horizontal and vertical dotted lines indicate the position and radial velocity of the AFGL2591 VLA 3 source. Panel {\bf h)} shows the combined map of the H$_2$S (2$_{2,0}$$\rightarrow$2$_{1,1}$), SO$_2$ (11$_{5,7}$$\rightarrow$12$_{4,8}$) and CH$_3$OH $v_t$=0 (4$_{+2,2}$$\rightarrow$3$_{+1,2}$ E) lines. The first contour and step levels for the combined map are 70 (2$\sigma$) and 105$\,$mJy/beam. Dashed, solid and dotted lines show the PV curves calculated assuming Keplerian rotation for a source with central mass 10$\,$M$_\odot$, 40$\,$M$_\odot$ and 100$\,$M$_\odot$, respectively. The envelope size and inclination angle assumed for these PV curves are 5400$\,$AU and i=20$^\circ$, respectively.}
\label{f5}
\end{center}
\end{figure*}

The presence of a velocity gradient for the molecular gas across AFGL2591 VLA 3 is also evident from the Position-Velocity (PV) diagrams obtained along the PV cuts shown in Figures$\,$\ref{f3} and \ref{f4} (see Figure$\,$\ref{f5}). The direction of the PV cuts (P.A.$\sim$20$^\circ$) has been selected so that it is roughly perpendicular to the axis of the AFGL2591 VLA 3 outflow \citep[P.A.$\sim$280$^\circ$;][]{van99}.
The offsets of the PV diagrams (in arcseconds) are calculated from the northeast to the southwest of AFGL2591 VLA 3. 

%FIG5**************************************
%******************************************

From Figure$\,$\ref{f5}, we find that Type I, II and III species show significant differences in their kinematic distribution with {\it holes} in the PV diagrams of SO$_2$, SO and CH$_3$OH. The kinematic structure of all these molecules, however, seems to be complementary, as expected if they probed different regions within the envelope. As shown in Figure$\,$\ref{f5}, H$_2$S (Type I) traces the regions closer to the VLA 3 protostar and is detected across a larger velocity range (from -12$\,$km$\,$s$^{-1}$ to 0$\,$km$\,$s$^{-1}$) than SO$_2$ (from -10$\,$km$\,$s$^{-1}$ to -2$\,$km$\,$s$^{-1}$) or CH$_3$OH (from -8$\,$km$\,$s$^{-1}$ to -3$\,$km$\,$s$^{-1}$). In contrast to H$_2$S, SO$_2$ (Type II species) peaks at regions further away (at $\sim$0.25$"$-0.35$"$) from the position of the central source, while CH$_3$OH (Type III) is found at $\sim$0.5$"$ away from VLA 3.

Since every species probes a particular region within the AFGL2591 VLA 3 hot core, the global gas kinematics of the envelope can be obtained by combining the data from Type I, Type II and Type III species. For this, we have used the molecular line transitions H$_2$S (2$_{2,0}$$\rightarrow$2$_{1,1}$), SO$_2$ (11$_{5,7}$$\rightarrow$12$_{4,8}$) and CH$_3$OH $v_t$=0 (4$_{+2,2}$$\rightarrow$3$_{+1,2}$ E). The combined image is generated by normalizing the individual line datacubes to their measured peak intensities and, after normalization, the individual images are added by considering that each of them accounts for 1/3 of the final image. The combined image is shown in panel {\bf h} of Figure$\,$\ref{f5}. We note that if H$_2$S were not included in the combined image, the final PV diagram would show a central hole in its distribution preventing us to characterize the very inner regions with the most blue- and red-shifted velocities toward the AFGL2591 VLA 3 hot core.

In Figure$\,$\ref{f5} (panel {\bf h}), we also compare the combined map with the expected PV diagrams from a rotating envelope that follows a Keplerian law around a central source with a mass of 10$\,$M$_\odot$, 40$\,$M$_\odot$ and 100$\,$M$_\odot$ (see dashed, solid and dotted lines, respectively). The kinematics of the molecular gas toward the AFGL2591 VLA 3 envelope can be explained by Keplerian-like rotation around a 40$\,$M$_\odot$-source. The derived envelope size is 5400$\,$AU, and the inclination angle is i=20$^\circ$. This angle is consistent with that proposed by \citet{van99} and \citet{wiel11} for this source. The derived mass of the central object, 40$\,$M$_\odot$, is consistent with that estimated by \citet[][of $\sim$38$\,$M$_\odot$]{san12}, and corresponds to an O6 star on the ZAMS. Previous estimates of the mass of the AFGL2591 VLA 3 source were underestimated due to the closer distance considered \citep[of $\sim$1$\,$kpc; see e.g.][]{van99,tri03}. 

\subsection{UV coverage and missing flux in the VEX images}
\label{uv}

%FIG6**************************************
\begin{figure}
\begin{center}
\includegraphics[angle=270,width=0.45\textwidth]{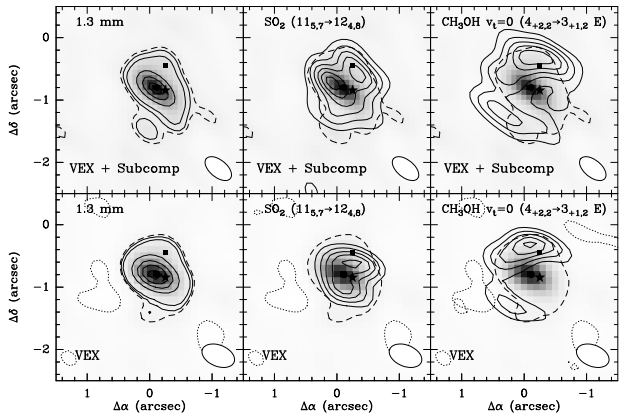}
\caption{Comparison of the combined (VEX + Subcompact) integrated intensity images of the 1.3$\,$mm continuum emission, and of the SO$_2$ (11$_{5,7}$$\rightarrow$12$_{4,8}$) and CH$_3$OH $v_t$=0 (4$_{+2,2}$$\rightarrow$3$_{+1,2}$ E) lines, with those obtained by using the VEX data only. The velocity range considered to generate the SO$_2$ and CH$_3$OH images goes from -8.7 to -2.1$\,$km$\,$s$^{-1}$. For the 1.3$\,$mm images (color and thin contours in left panels), the contour levels are 6.0 (4$\sigma$), 7.5, 10.5, 25.5, 40.5 and 55.5$\,$mJy/beam for the VEX + Subcompact data, and of 3.2 (4$\sigma$), 4.0, 5.6, 17.6, 29.6, 41.6 and 53.6$\,$mJy/beam for the VEX map. The SO$_2$ and CH$_3$OH emission maps (thick contours in central and right panels) appear superimposed on the 1.3$\,$mm continuum images (color scale and dashed contour). For the SO$_2$ and CH$_3$OH line data, the first contour and step levels are, respectively, 0.5 (5$\sigma$) and 0.3$\,$Jy/beam$\,$km$\,$s$^{-1}$ for the combined VEX + Subcompact images, and 0.35 (5$\sigma$) and 0.21$\,$Jy/beam$\,$km$\,$s$^{-1}$ for the VEX maps. Symbols are as in Figures$\,$\ref{f3} and \ref{f4}. Beams are shown at the lower right corner in every panel.}
\label{f6}
\end{center}
\end{figure}
%******************************************

The differences in the morphology between Type I, II and III species toward AFGL2591 cannot be attributed to differences in the UV plane coverage, because all molecular lines were observed simultaneously within the same track (see Figure$\,$\ref{f1}). 

Alternatively, one may think that the observed morphological differences are due to missing flux in the VEX observations. In Figure$\,$\ref{f6}, we compare the combined images of the 1.3$\,$mm continuum, SO$_2$ (Type II) and CH$_3$OH emission (Type III species) obtained with the VEX + Subcompact data, with those generated with the VEX data only. From Figure$\,$\ref{f6}, we find that the combined VEX + Subcompact maps 
provide a very similar spatial distribution to that reported in the VEX images. This demonstrates that the observed differences in the morphology of Type I, II and III species are not produced by missing short spacings, but due to a real chemical segregation within the AFGL2591 VLA 3 hot core. We note that in Figure$\,$\ref{f6} we do not include the combined image of the H$_2$S (2$_{2,0}$$\rightarrow$2$_{1,1}$) transition, because this line was not covered within the Subcompact observations (see Section$\,$\ref{obs}).

\section{Excitation temperature, column densities and molecular abundances toward AFGL2591}
\label{exc}

From the integrated intensities of the SO$_2$, CH$_3$OH and HC$_3$N lines measured toward single beams across the AFGL2591 VLA 3 hot core, we can estimate the excitation temperature of the gas across the envelope, T$_{ex}$, by means of the population diagram method that assumes optically thin emission and LTE \citep{gol99}. 
We select the Continuum Peak position, and the emission peaks T1 and T2 because they show the brightest molecular emission for Type I, II and III species, respectively (see Figure$\,$\ref{f3}). In Table$\,$\ref{tab2}, we report the observed parameters of the single beam lines measured toward the Continuum Peak for H$_2$S and $^{13}$CS, and toward emission peaks T1 for SO, OCS, HC$_3$N and SO$_2$, and T2 for CH$_3$OH (see Figure$\,$\ref{f3} and this Table for the exact coordinates of these positions). The derived central radial velocities, $v_{LSR}$, and linewidths, $\Delta v$, typically range from -5.5 to -6.5$\,$km$\,$s$^{-1}$, and from $\sim$2.5 to 4$\,$km$\,$s$^{-1}$, respectively, although significant discrepancies in $v_{LSR}$ and $\Delta v$ are found for some of the species within the same type (see e.g. SO for Type II or the low-excitation lines CH$_3$OH $v_t$=0 (5$_{+1,4}$$\rightarrow$4$_{+2,2}$ E) and CH$_3$OH $v_t$=0 (3$_{-2,2}$$\rightarrow$4$_{-1,4}$ E) for Type III; Table$\,$\ref{tab2}). These discrepancies could be due to large optical depths associated with these low-excitation lines ($E_u$$\leq$100$\,$K; see Table$\,$\ref{tab1}).

\begin{figure*}
\begin{center}
\includegraphics[angle=270,width=0.85\textwidth]{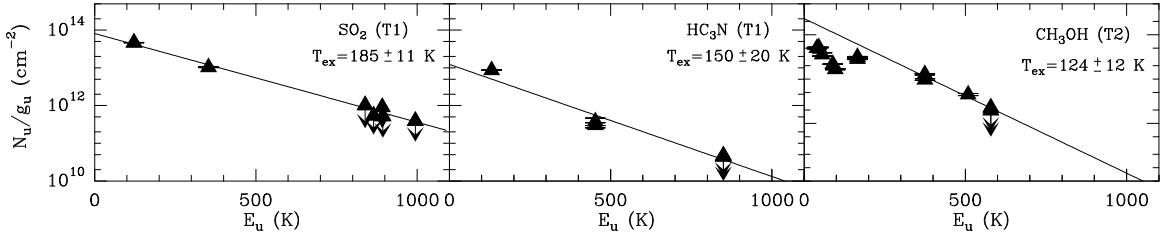}
\caption{Population diagrams for SO$_2$, HC$_3$N and CH$_3$OH derived from the line spectra extracted from positions T1 and T2 toward AFGL2591 VLA 3 (see Table$\,$\ref{tab2}). The derived T$_{ex}$ are shown in the upper right corner of every panel. Error bars correspond to 1$\sigma$ uncertainties and vertical arrows indicate upper limits.}
\label{f7}
\end{center}
\end{figure*}

In Figure$\,$\ref{f7}, we show the population diagrams of SO$_2$ and HC$_3$N derived toward position T1, and the population diagram of CH$_3$OH obtained toward position T2. The derived T$_{ex}$ are 185$\pm$11$\,$K for SO$_2$, 150$\pm$20$\,$K for HC$_3$N, and 124$\pm$12$\,$K for CH$_3$OH, suggesting a temperature gradient within the envelope. The derived column densities of SO$_2$, HC$_3$N and CH$_3$OH are 2$\times$10$^{17}$$\,$cm$^{-2}$, 8$\times$10$^{15}$$\,$cm$^{-2}$ and 4$\times$10$^{17}$$\,$cm$^{-2}$, respectively. In the case of CH$_3$OH, T$_{ex}$ was derived from lines with $E_u$$\geq$300$\,$K because its lower-excitation transitions are likely optically thick. As shown by \citet{gir02}, large optical depths indeed lead to the underestimation of the measured column densities of the low-excitation transitions of CH$_3$OH, clearly deviating from the optically thin solution provided by the population diagram method.

%TABLE2------------------------------------------------------------------
\begin{deluxetable*}{lcccc}
\tabletypesize{\scriptsize}
\tablecaption{Observed parameters for the molecular lines measured
toward the continuum peak, and toward positions T1 and T2 in AFGL2591 VLA 3.}
\tablewidth{0pt}

\startdata

%\hline \hline

& & TABLE 2 & & \\ \hline \hline
Molecule & Transition & \multicolumn{3}{c}{{\bf Continuum Peak\tablenotemark{a}}} \\ \cline{3-5}
& & $v_{LSR}$ (km$\,$s$^{-1}$)& $\Delta v$ (km$\,$s$^{-1}$) & Peak Intensity (Jy/beam) \\ \hline
H$_2$S & 2$_{2,0}$$\rightarrow$2$_{1,1}$ & -5.4(0.5) & 7.3(1.2) & 0.34(0.03) \\
$^{13}$CS & 5$\rightarrow$4 & -5.44(0.16) & 3.7(0.5) & 0.18(0.02) \\ \hline 
& & \multicolumn{3}{c}{{\bf Position T1\tablenotemark{b}}} \\ \cline{3-5}
& & $v_{LSR}$ (km$\,$s$^{-1}$)& $\Delta v$ (km$\,$s$^{-1}$) & Peak Intensity (Jy/beam) \\ \hline
SO & 5$_6$$\rightarrow$4$_5$ & -4.71(0.06) & 6.42(0.13) & 0.60(0.02) \\
OCS & 18$\rightarrow$17 & -6.27(0.09) & 3.6(0.2) & 0.352(0.016) \\
& 19$\rightarrow$18 & -6.21(0.16) & 4.5(0.4) & 0.21(0.02) \\ 
HC$_3$N & $v$=0 24$\rightarrow$23 & -6.8(0.1) & 3.7(0.2) & 0.28(0.02) \\
& $v_7$=1$_e$ 24$\rightarrow$23 & -7.2(0.4) & 1.3(0.0) & 0.10(0.03) \\
& $v_7$=1$_f$ 24$\rightarrow$23 & -7.7(0.4) & 1.7(0.5) & 0.06(0.02) \\
& $v_6$=1$_e$ 24$\rightarrow$23 & $\dots$ & $\ldots$ & $\leq$0.06\tablenotemark{d} \\
& $v_6$=1$_f$ 24$\rightarrow$23 & $\dots$ & $\ldots$ & $\leq$0.05 \\
SO$_2$ & 22$_{7,15}$$\rightarrow$23$_{6,18}$ & -5.81(0.12) & 3.1(0.3) & 0.193(0.017) \\
& 22$_{2,20}$$\rightarrow$22$_{1,21}$ & $\dots$ & $\ldots$ & $\leq$0.05 \\
& 16$_{3,13}$$\rightarrow$16$_{2,14}$ & $\dots$ & $\ldots$ & $\leq$0.05 \\
& 11$_{5,7}$$\rightarrow$12$_{4,8}$ & -5.69(0.07) & 3.41(0.18) & 0.356(0.018) \\
& 13$_{2,12}$$\rightarrow$13$_{1,13}$ & $\dots$ & $\ldots$ & $\leq$0.06 \\
& 37$_{10,28}$$\rightarrow$38$_{9,29}$ & $\dots$ & $\ldots$ & $\leq$0.05 \\
& 14$_{3,11}$$\rightarrow$14$_{2,12}$ & $\dots$ & $\ldots$ & $\leq$0.05 \\ \hline
& & \multicolumn{3}{c}{{\bf Position T2\tablenotemark{c}}} \\ \cline{3-5}
& & $v_{LSR}$ (km$\,$s$^{-1}$)& $\Delta v$ (km$\,$s$^{-1}$) & Peak Intensity (Jy/beam) \\ \hline
CH$_3$OH & $v_t$=0 5$_{+1,4}$$\rightarrow$4$_{+2,2}$ $E$ & -5.3(0.5) & 4.1(1.0) & 0.07(0.03) \\
& $v_t$=1 6$_{1,5}$$\rightarrow$7$_{2,6}$ $A^-$ & -6.41(0.13) & 2.4(0.3) & 0.174(0.018) \\
& $v_t$=0 20$_{+1,19}$$\rightarrow$20$_{+0,20}$ $E$ & -5.95(0.15) & 2.1(0.4) & 0.18(0.02) \\
& $v_t$=0 4$_{+2,2}$$\rightarrow$3$_{+1,2}$ $E$ & -5.93(0.07) & 3.29(0.15) & 0.39(0.02) \\
& $v_t$=0 8$_{+0,8}$$\rightarrow$7$_{+1,6}$ $E$ & -5.7(0.3) & 4.9(0.7) & 0.087(0.017) \\
& $v_t$=0 15$_{+4,11}$$\rightarrow$16$_{+3,13}$ $E$ & -6.0(0.3) & 4.4(0.8) & 0.085(0.018) \\
& $v_t$=0 8$_{-1,8}$$\rightarrow$7$_{+0,7}$ $E$ & -5.74(0.08) & 3.14(0.18) & 0.290(0.018) \\
& $v_t$=0 19$_{5,15}$$\rightarrow$20$_{4,16}$ $A^+$ & $\dots$ & $\ldots$ & $\leq$0.05 \\
& $v_t$=0 19$_{5,14}$$\rightarrow$20$_{4,17}$ $A^-$ & $\dots$ & $\ldots$ & $\leq$0.04 \\
& $v_t$=0 3$_{-2,2}$$\rightarrow$4$_{-1,4}$ $E$ & -7.13(0.15) & 2.5(0.4) & 0.124(0.016) \\
& $v_t$=0 10$_{2,9}$$\rightarrow$9$_{3,6}$ $A^-$ & -6.63(0.17) & 3.8(0.4) & 0.19(0.02) \\
& $v_t$=0 10$_{2,8}$$\rightarrow$9$_{3,7}$ $A^+$ & -6.3(0.2) & 4.1(0.5) & 0.16(0.02) \\ \hline

\enddata

\tablenotetext{a}{The coordinates of the Continuum Peak are $\alpha (J2000)$=20$^h$29$^m$24.889$^s$, $\delta (J2000)$=40$^\circ$11$'$19.51$"$.}
\tablenotetext{b}{The coordinates for the T1 position are $\alpha (J2000)$=20$^h$29$^m$24.870$^s$, $\delta (J2000)$=40$^\circ$11$'$19.73$"$.}
\tablenotetext{c}{The coordinates for position T2 are $\alpha (J2000)$=20$^h$29$^m$24.884$^s$, $\delta (J2000)$=40$^\circ$11$'$20.00$"$.}
\tablenotetext{d}{Upper limits refer to the 3$\sigma$ level in the extracted single beam line spectra.}

\label{tab2}
\end{deluxetable*}
%------------------------------------------------------------------

For H$_2$S, $^{13}$CS, OCS, and SO, the molecular column densities are estimated from single transitions by considering optically thin emission and LTE conditions. The assumed excitation temperature for all these lines is T$_{ex}$=185$\,$K, the excitation temperature found for SO$_2$ toward T1. For $^{13}$CS, we have also considered an isotopic ratio $^{12}$C/$^{13}$C=60 \citep{wil94}. The derived column densities are 7$\times$10$^{16}$$\,$cm$^{-2}$ for H$_2$S, 4$\times$10$^{14}$$\,$cm$^{-2}$ for $^{13}$CS, 2$\times$10$^{16}$$\,$cm$^{-2}$ for OCS, and 3$\times$10$^{16}$$\,$cm$^{-2}$ for SO.

Since H$_2$S and $^{13}$CS are mainly detected toward the Continuum Peak, it is possible that their excitation temperature is higher than assumed here, as suggested by our modelling of the chemical segregation toward AFGL2591 VLA 3 (see Section$\,$\ref{chem}). In case T$_{ex}$ were $\sim$1000$\,$K \citep[as measured by][from ro-vibrational absorption lines of HCN]{carr95}, the column densities derived for H$_2$S and $^{13}$CS would be factors $\sim$5-8 higher than those reported above.

To estimate the molecular abundances across the AFGL2591 VLA 3 envelope, we used the 1.3$\,$mm continuum flux toward the Continuum Peak (68$\,$mJy/beam), toward T1 (19$\,$mJy/beam), and toward T2 (8$\,$mJy/beam), to derive the H$_2$ column densities at these positions. From Equation$\,$2 in \citet{eno06}, and by considering dust opacities of 0.01$\,$cm$^2$$\,$g$^{-1}$ \citep{oss94}, we estimate H$_2$ column densities of 2$\times$10$^{24}$$\,$cm$^{-2}$, 4$\times$10$^{23}$$\,$cm$^{-2}$ and 3$\times$10$^{23}$$\,$cm$^{-2}$ toward the Continuum Peak, T1 and T2, respectively. We have assumed dust temperatures of $\sim$200$\,$K toward the Continuum Peak, of $\sim$200$\,$K toward T1, and of $\sim$120$\,$K toward T2, following the gas excitation temperatures derived from SO$_2$ and CH$_3$OH (see above). However, we note that these H$_2$ column densities are subject to uncertainties since the dust temperature cannot be constrained from our observations. If the temperature of the dust were higher than 200$\,$K toward the Continuum Peak as suggested by the modeling of \citet[][$\sim$500-800$\,$K for angular scales $\leq$0.2$"$]{van99}, the derived H$_2$ column densities would be decreased to $\sim$3-5$\times$10$^{23}$$\,$cm$^{-2}$. 

The derived molecular abundances are presented in Columns 2 and 4 of Table$\,$\ref{tab3}. In this Table, we also include the HCN abundances measured toward AFGL2591 and reported by \citet{doty02}. HCN is detected in absorption in the IR toward this source \citep{carr95}; and since absorption is restricted to the direction of the line-of-sight toward the source, the bulk of the HCN gas is likely associated with the innermost regions of the AFGL2591 VLA 3 envelope. We do not include the abundance from other molecular species reported in \citet{doty02}, because they are mainly obtained from emission measurements toward this source, which likely probe gas at all scales throughout the envelope. 

From Table$\,$\ref{tab3}, we find that the derived abundances decrease with increasing radius from the position of the Continuum peak to positions T1 and T2 by factors of $\geq$5-10 for H$_2$S and CS, and by two orders of magnitude for HCN. For HC$_3$N, SO$_2$, OCS and SO, however, their derived abundances are factors $\geq$6 larger toward position T1 than toward the Continuum Peak. Following a similar trend, the derived abundance of CH$_3$OH is a factor of $\geq$100 larger toward position T2 than toward the Continuum Peak. 

We note that, if the assumed excitation temperature of the gas were $\sim$800$\,$K toward the continuum peak, the measured molecular abundances toward this position (see Column 2 in Table$\,$\ref{tab3}) would be increased by factors of $\sim$2-40. This implies that the increasing abundance trend measured for Type II species (i.e. HC$_3$N, SO$_2$, OCS and SO) across AFGL2591 VLA 3 would become non-existent. 
However, in the case of CH$_3$OH (Type III species), the derived upper limit to its abundance ($\leq$2$\times$10$^{-7}$) would still be one order of magnitude lower than the CH$_3$OH abundance measured toward position T2 (Table$\,$\ref{tab3}); and for Type I species (H$_2$S and CS), their decreasing abundance trend with larger radii would be even stronger than reported in Table$\,$\ref{tab3} (the H$_2$S and CS abundances measured toward the Continuum Peak would be factors of $\sim$100-500 higher than toward T2).

We finally stress that the scope of this paper is to show the abundance trends found across the AFGL2591 VLA 3 envelope, and not to obtain the global molecular abundances measured toward this object for which observations with lower angular resolution would be required. 

%FIG7**************************************
%******************************************

%TABLE3------------------------------------------------------------------
\begin{deluxetable*}{lcccc}
\tabletypesize{\scriptsize}
\tablecaption{Comparison of the observed molecular abundances with those predicted by our two-point model of the AFGL2591 VLA 3 envelope.}
\tablewidth{0pt}

\startdata

%\hline \hline

& & TABLE 3 & & \\ \hline \hline
Abundances & Observed & Model\tablenotemark{a} & Observed & Model\tablenotemark{a} \\
& {\bf (Continuum Peak)} & {\bf (Region A)} & {\bf (Positions T1/T2)} & {\bf (Region B)} \\ \hline
H$_2$S & 5$\times$10$^{-8}$ & 5$\times$10$^{-9}$ & $\leq$4$\times$10$^{-9}$\tablenotemark{b} & 1$\times$10$^{-10}$ \\
CS & 2$\times$10$^{-8}$ & 1$\times$10$^{-6}$ & $\leq$4$\times$10$^{-9}$\tablenotemark{b} & 1$\times$10$^{-7}$ \\
HCN & 1$\times$10$^{-6}$\tablenotemark{c} & 4$\times$10$^{-5}$ & 1$\times$10$^{-8}$\tablenotemark{c} & 3$\times$10$^{-7}$ \\ 
HC$_3$N & 7$\times$10$^{-9}$ & 4$\times$10$^{-11}$ & 2$\times$10$^{-8}$ & 3$\times$10$^{-8}$ \\
OCS & 9$\times$10$^{-9}$ & 1$\times$10$^{-11}$ & 6$\times$10$^{-8}$ & 2$\times$10$^{-9}$ \\
SO$_2$ & 1$\times$10$^{-7}$ & 3$\times$10$^{-11}$ & 6$\times$10$^{-7}$ & 1$\times$10$^{-6}$ \\
SO & 1$\times$10$^{-8}$ & 2$\times$10$^{-9}$ & 6$\times$10$^{-8}$ & 1$\times$10$^{-8}$ \\
CH$_3$OH & $\leq$2$\times$10$^{-8}$\tablenotemark{d} & $\leq$1$\times$10$^{-13}$ & 2$\times$10$^{-6}$\tablenotemark{e} & 5$\times$10$^{-8}$ \\ \hline

\enddata

\tablenotetext{a}{Abundances predicted by the model for a dynamical age of $\sim$2$\times$10$^4$$\,$yr \citep{doty02}.}
\tablenotetext{b}{Upper limits derived from the 1$\sigma$ level in the H$_2$S and $^{13}$CS spectra toward position T2.}
\tablenotetext{c}{From \citet{doty02}.}
\tablenotetext{d}{Upper limits derived from the 1$\sigma$ level in the CH$_3$OH spectrum toward the position of the Continuum Peak.}
\tablenotetext{e}{CH$_3$OH abundance derived from the population diagram shown in Figure$\,$\ref{f7} for position T2.}

\label{tab3}
\end{deluxetable*}
%------------------------------------------------------------------

\section{Chemical modeling of the AFGL2591 hot core}
\label{chem}

To qualitatively explain the chemical segregation observed toward AFGL2591 VLA 3, we have used the two-phase UCL CHEM gas-grain chemical code \citep{viti04}. In Phase I, this code simulates the free-fall gravitational collapse of a cloud where the mantles of dust grains form via hydrogenation reactions. The initial atomic abundances are solar \citep{sof01}. The formation of the initial core occurs from a diffuse medium with H$_2$ densities of $\sim$100$\,$cm$^{-3}$ at a kinetic temperature of 20$\,$K. The core collapse is stopped when the final density is reached in the model, and it corresponds to time-scales of $\sim$5$\times$10$^6$$\,$yrs. In Phase 2, we calculate the time-dependent evolution of the chemistry of gas and dust once stellar activity is present. In our case, we assume that the mantles are instantaneously evaporated after the {\it turning on} of the protostar. We also assume that H$_2$S is the main reservoir of atomic sulfur on dust grains.

The detection of ro-vibrational absorption of HCN toward AFGL2591 VLA 3 \citep{carr95,lau97}, suggests the presence of an inner and hotter core with relatively low extinction ($A_v$$\leq$30$^m$), embedded within a more extended and cooler envelope \citep[][]{van99}. The AFGL2591 VLA 3 hot core has therefore been modeled taking into account these two physical regimes (regions A and B; Table$\,$\ref{tab4}). Region A corresponds to the inner core where the emission from H$_2$S and $^{13}$CS peaks (i.e. toward the Continuum Peak with angular scales of $\sim$0.2$"$ or $r$=600$\,$AU at a distance of 3$\,$kpc). Region B corresponds to an intermediate position between T1 and T2 in our observations where HC$_3$N, OCS, SO$_2$, SO and CH$_3$OH show their maximum emission (i.e. at $\sim$0.4$"$, or $r$=1100$\,$AU, from the central source in our images; see Figure$\,$\ref{f3}). Therefore, for simplicity, our two region model does not discriminate between positions T1 and T2 within the AFGL2591 VLA 3 envelope. We also consider the existence of an inner cavity within the AFGL2591 VLA 3 hot core, as proposed by \citet{prei03} and \citet{dewit09}. These authors derived a radius of 40$\,$AU for this cavity assuming a distance of $\sim$1$\,$kpc for the AFGL2591 VLA 3 source \citep{van99}. Since the actual distance to this source is 3$\,$kpc \citep{rygl12}, the radius of the cavity considered in our model is 120$\,$AU, i.e. a factor of 3 larger than that derived by \citet{prei03} or \citet{dewit09}. This cavity remains unresolved in our VEX images. The wall of the cavity at $r$=120$\,$AU is the position at which the visual extinction in our model is $A_v$=0.

The H$_2$ densities at the end of Phase I for regions A and B are estimated by using the H$_2$ density distribution calculated by \citet{van00a}, after correcting it from the distance of 3$\,$kpc \citep[see discussion in Section$\,$4.2 of][on how the H$_2$ density distribution for AFGL2591 VLA 3 is affected by a larger distance to the source]{van99}. The averaged H$_2$ densities for regions A and B are, respectively, 4.6$\times$10$^6$$\,$cm$^{-3}$ and 1.8$\times$10$^6$$\,$cm$^{-3}$. These densities are kept constant in Phase II of our model. 

In order to compare the H$_2$ column densities expected from these H$_2$ density values with those measured toward the Continuum Peak and positions T1/T2, one needs to consider the total H$_2$ column density across the AFGL2591 VLA 3 envelope. For instance, to compare the results from region A with those observed toward the Continuum Peak, the contribution from the inner core ($\sim$5$\times$10$^6$$\,$cm$^{-3}$$\times$960$\,$AU, i.e. 2$\times$480$\,$AU, the shell size for region A; see below) must be added to that from the outer envelope [$\sim$2$\times$10$^6$$\,$cm$^{-3}$$\times$(5400-960)$\,$AU, where 5400$\,$AU is the envelope size derived in Section$\,$\ref{kin}]. This gives a total H$_2$ column density of $\sim$2$\times$10$^{23}$$\,$cm$^{-2}$. For region B (or positions T1/T2), the total H$_2$ column density can be estimated simply by doing $\sim$2$\times$10$^6$$\,$cm$^{-3}$$\times$5400$\,$AU, which gives $\sim$1$\times$10$^{23}$$\,$cm$^{-2}$. For region B, the expected and observed H$_2$ column densities are within a factor of $\sim$4. However, for region A (or the Continuum Peak), the low value of the expected H$_2$ column density (2$\times$10$^{23}$$\,$cm$^{-2}$ vs. 2$\times$10$^{24}$$\,$cm$^{-2}$; see Section$\,$\ref{exc}) suggests that the actual temperature of the dust toward the Continuum Peak is likely higher than 200$\,$K (Section$\,$\ref{exc}).

The gas and dust temperatures assumed for regions A and B are obtained from the theoretical modelling of the core by \citet{sta04} for region A ($\sim$1000$\,$K), and by \citet{van99} for region B ($\sim$200$\,$K). We stress that the dust and gas temperature profiles do not need to be corrected by the distance of 3$\,$kpc because they are constant in terms of projected distance \citep[i.e. in arcseconds; see Section$\,$4.2 of][]{van99}. 

%FIG8**************************************
\begin{figure}
\begin{center}
\includegraphics[angle=270,width=0.5\textwidth]{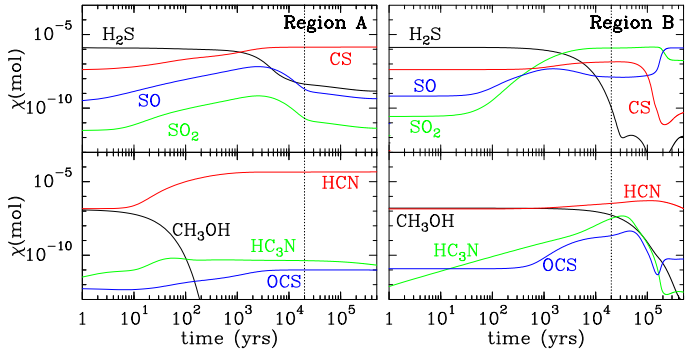}
\caption{Abundances of H$_2$S, CS, SO$_2$, SO, HCN, HC$_3$N, OCS and CH$_3$OH predicted for Phase 2 by the two-point chemical model of the AFGL2591 VLA 3 envelope (Section$\,$\ref{chem}). The physical properties assumed for regions A and B are reported in Table$\,$\ref{tab4}. Vertical dotted lines show the dynamical age of the AFGL2591 envelope ($\sim$2$\times$10$^4$$\,$yr) derived by \citet{doty02} and \citet{sta05}.}
\label{f8}
\end{center}
\end{figure}
%******************************************

%TABLE4------------------------------------------------------------------
\begin{deluxetable}{lcccc}
\tabletypesize{\scriptsize}
\tablecaption{Physical parameters of the two-point chemical model of the AFGL2591 VLA 3 envelope}
\tablewidth{0pt}

\startdata

%\hline \hline

& & TABLE 4 & & \\ \hline \hline
& \multicolumn{2}{c}{{\bf Region A}} & \multicolumn{2}{c}{{\bf Region B}} \\ \hline
$r$ & \multicolumn{2}{c}{600 AU} & \multicolumn{2}{c}{1100 AU} \\ 
$n$(H$_2$)\tablenotemark{a} & \multicolumn{2}{c}{4.6$\times$10$^6$ cm$^{-3}$} & \multicolumn{2}{c}{1.8$\times$10$^6$ cm$^{-3}$} \\
T$_{kin}$\tablenotemark{b} & \multicolumn{2}{c}{1000 K} & \multicolumn{2}{c}{200 K} \\ 
Av\tablenotemark{c} & \multicolumn{2}{c}{20.5$^{m}$} & \multicolumn{2}{c}{28.5$^{m}$} \\ \hline

\enddata

\tablenotetext{a}{From \citet{boon01}.}
\tablenotetext{b}{Consistent with HCN line absorption data \citep{carr95,van99} and with our SO$_2$ data (Section$\,$\ref{exc}).}
\tablenotetext{c}{See text for the derivation of $A_v$.}
%\tablenotetext{d}{The effective radius used in the model was $r_{eff}$=1650$\,$AU to obtain an extinction of 28.5$^{m}$ (see text).}

\label{tab4}
\end{deluxetable}
%------------------------------------------------------------------

To derive the FUV luminosity emitted by the AFGL2591 VLA 3 source, we have used the FUV luminosity profile derived by \citet[][]{bru09a} as a function of effective temperature, T$_{eff}$ (see Figure$\,$A.1 in their work). For an O6-type star on the ZAMS with an effective temperature of T$_{eff}$$\sim$4$\times$10$^4$$\,$K, the luminosity in the FUV band is $\sim$2$\times$10$^{33}$$\,$erg$\,$s$^{-1}$ per L$_\odot$. Considering that the AFGL2591 VLA 3 source has a total bolometric luminosity of L$_{bol}$$\sim$2$\times$10$^5$$\,$L$_\odot$ \citep[Section$\,$\ref{res} and][]{san12}, we obtain that the total FUV luminosity emitted by this source is 4$\times$10$^{38}$$\,$erg$\,$s$^{-1}$. This luminosity is a factor of 10 higher than that used by \citet{bru09a}, because the distance assumed in their calculations was 1$\,$kpc instead of 3$\,$kpc. Since our model considers a cavity within the AFGL2591 VLA 3 envelope, the FUV photon flux illuminating the wall of this cavity (at 120$\,$AU) is $\sim$6$\times$10$^9$$\,$hab \citep[1$\,$hab=1.6$\times$10$^{-3}$$\,$erg$\,$s$^{-1}$$\,$cm$^{-2}$;][]{hab68}, as derived from geometrical dilution. We thus assume an incident FUV flux of $\sim$6$\times$10$^9$$\,$hab at the position with $A_v$=0 in our model.

At $r$$>$120$\,$AU, the extinction is calculated as $A_v$=[$size$$\times$$n$(H$_2$)]/[1.6$\times$10$^{21}$$\,$cm$^{-2}$]. For region A (shell thickness $size$=600$\,$AU-120$\,$AU=480$\,$AU), the extinction is $\sim$20.5$^{m}$. For region B, however, the individual extinction from regions A and B should be added since the FUV field from the AFGL2591 star will be attenuated as it propagates outwards through the envelope. The individual extinction at $r$=1100$\,$AU (shell thickness $size$$\sim$1100$\,$AU-600$\,$AU=500$\,$AU) is $\sim$8$^{m}$, leading to a total extinction of $\sim$28.5$^m$ (20.5$^m$+8$^m$) for region B. 

In Figure$\,$\ref{f8}, we present the abundances of H$_2$S, CS, HCN, SO, SO$_2$, HC$_3$N, OCS and CH$_3$OH, predicted for Phase 2 by our two-point model of the AFGL2591 VLA 3 envelope. Our results show that the chemical segregation observed toward this source can be explained by the combination of molecular UV photo-dissociation and a high-temperature gas-phase chemistry in the inner core of AFGL2591 VLA 3. In region A, H$_2$S and CH$_3$OH are initially destroyed by FUV photons (extincted FUV flux of $\sim$7.5$\,$hab for $A_v$$\sim$20.5$^m$) after the release of the grain mantles into the gas-phase ($t$$\leq$100$\,$yr). However, the high temperatures of the gas (T$_{kin}$$\sim$1000$\,$K) allow the re-formation of H$_2$S via the endothermic reaction H$_2$$\,$+$\,$HS$\rightarrow$H$_2$S$\,$+$\,$H (activation barrier of 8050$\,$K), making H$_2$S abundant in the inner core (5$\times$10$^{-9}$ for t$\sim$2$\times$10$^4$$\,$yr; Figure$\,$\ref{f8}). We note that these high temperatures are strictly required for region A because the H$_2$S abundance would dramatically drop otherwise, in contrast with what is observed. Like H$_2$S, HCN and CS survive the FUV photo-dissociation thanks to the gas-phase reactions H$_2$$\,$+$\,$CN$\rightarrow$HCN$\,$+$\,$H and C$_2$$\,$+$\,$S$\rightarrow$CS$\,$+$\,$C. CH$_3$OH is completely destroyed since it cannot be re-formed in the gas phase \citep[e.g.][]{garr08}. 

In region B, molecular photo-dissociation is less severe (extincted FUV flux of only $\sim$0.003$\,$hab for $A_v$$\sim$28.5$^m$); and since the gas temperature is only 200$\,$K, the activation barrier for the endothermic reaction to form H$_2$S cannot be overcome. As a consequence, other S-bearing products such as SO, OCS and SO$_2$ are enhanced (Figure$\,$\ref{f8}). This explains the detection of H$_2$S only toward the inner core of AFGL2591 and the double-peaked structure observed for SO, SO$_2$ and OCS (Section$\,$\ref{res}). We note that this S-bearing chemical structure was also predicted by more complex chemical modeling of the AFGL2591 VLA 3 source, which includes UV and X-ray photo-chemistry \citep{bru09b}. 

In the case of CH$_3$OH and HC$_3$N, these molecules are found to survive in region B thanks to the higher extinction. For HCN, the formation of this molecule is less efficient at lower temperatures, explaining the decrease of the HCN abundance (by a factor of $\geq$100) in region B (Figure$\,$\ref{f8}). This decrease is consistent with that observed toward the AFGL2591 VLA 3 envelope \citep[Table$\,$\ref{tab3};][]{van99,boon01}. 

The agreement between the predicted and measured abundances for the rest of the molecular species is within factors $\sim$2-50, except for HC$_3$N, OCS and SO$_2$ for region A where the discrepancies are found to be of several orders of magnitude. This could be due to the simplistic physical structure assumed for the AFGL2591 VLA 3 envelope. Continuous variations of the density and temperature of the gas and dust with increasing radius, as derived by comprehensive radiative transfer and chemical models \citep[see e.g.][]{van99,doty02,bru09b}, could significantly change the predicted abundances of these molecules. Indeed, the molecular abundances are very sensitive to the extinction assumed for regions A and B ($A_v$ in Table$\,$\ref{tab4}), which depends on the H$_2$ density distribution of the envelope and the distance to the central protostar. If the assumed extinction is decreased by 1$^m$ for regions A and B, the general trend for the molecular abundance is to decrease by factors of $\sim$3-7, except for HC$_3$N, H$_2$S and OCS in region B where their abundances are decreased by factors $\geq$20. 

Alternatively, the large discrepancies in the measured and predicted abundances of HC$_3$N, OCS and SO$_2$ for region A could be due to the fact that the beam size of our VEX observations is larger (0.3$"$$\times$0.5$"$) than the angular scale considered for region A (of only 0.2$"$). The measured abundances for HC$_3$N, OCS and SO$_2$ could therefore have received some contribution from emission outside the inner 0.2$"$ of the envelope, artificially increasing the abundances for these molecules toward this region.

We note that the maximum extinction allowed for region B is 28.5$^m$. If an extinction of $A_v$$\geq$29$^m$ is assumed for this region, the abundance of H$_2$S increases to $\sim$10$^{-6}$, which is several orders of magnitude higher than found from our observations (the upper limit to the H$_2$S abundance measured toward position T2 is of $\leq$4$\times$10$^{-9}$; see Table$\,$\ref{tab3}).

Finally, our results do not depend on our selection of H$_2$S as the main carrier of S on the grain mantles. We have run two models with i) 100\% of S locked onto OCS; and ii) 50\% of S locked on H$_2$S and 50\% just as S. For region A, once OCS is released from the mantles, it is photo-dissociated leading to large abundances of S. S subsequently reacts to form HS, which is rapidly converted into H$_2$S. This also applies to the case with 50\% S and 50\% H$_2$S, where all S is transformed into H$_2$S. For region B, the only difference is found for the 100\% OCS model, where OCS is the most abundant molecule after the release of the mantles. OCS is then efficiently converted into SO and SO$_2$, showing a similar behavior to that found for our previous 100\% H$_2$S model. 

\section{Conclusions}
\label{conc}

We report the detection of a clear chemical segregation at linear scales $\leq$3000$\,$AU within the massive hot core around the high-mass protostar AFGL2591 VLA3 with the SMA. The molecular gas is distributed in three concentric shells where: i) H$_2$S and $^{13}$CS (Type I species) peak at the AFGL2591 VLA 3 continuum source; ii) HC$_3$N, OCS, SO$_2$ and SO (Type II species) show a double-peaked morphology circumventing the dust continuum peak; and iii) CH$_3$OH (Type III species) is distributed in a ring-like structure that surrounds the dust continuum emission. The global kinematics of the gas, probed by the different molecules found across the envelope, are consistent with Keplerian-like rotation around a central source of 40$\,$M$_\odot$. This mass is consistent with previous findings \citep{san12}. In addition, the molecular gas shows a temperature gradient for increasing distance to the central source, as revealed by the higher T$_{ex}$ derived for SO$_2$ (of 185$\pm$11$\,$K) than for CH$_3$OH (of 124$\pm$12$\,$K). By comparing our SMA images with gas and dust chemical modeling, we find that the chemical segregation toward the AFGL2591 VLA 3 hot core is explained by the combination of molecular UV photo-dissociation and a high-temperature gas-phase chemistry within the low extinction innermost region in the AFGL2591 VLA 3 massive hot core.    

\acknowledgments

We acknowledge the constructive comments from a second referee which helped to improve the paper. IJ-S acknowledges the Smithsonian Astrophysical Observatory for the support provided through a SMA fellowship. JM-P and IJ-S have been partially funded by MICINN grants ESP2007-65812-C02-C01, AYA2010-21697-C05-01 and AstroMadrid (CAM S2009/ESP-1496).

\clearpage

\end{document}